\documentclass[12pt]{article}
\usepackage{amsfonts}
\usepackage{epsfig}%
\usepackage{amsmath}%
\usepackage{amsfonts}%
\usepackage{amssymb}%
\usepackage{graphicx}

\usepackage[perpage,symbol*]{footmisc}

\def\addcite#1{[???]}
\def\chisb{\raise0.2em\hbox{$\chi$}SB}
\def\chiSM{\raise0.17em\hbox{$\chi$}SM}

\def\dslash{\not{\hbox{\kern-2pt $\partial$}}}
\def\Dslash{\not{\hbox{\kern-4pt  D }}}
\def\Aslash{\not{\hbox{\kern-4pt  A }}}
\def\TDslash{\not{\hbox{\kern-4pt $\tilde D$}}}
\def\Tdslash{\not{\hbox{\kern-2pt $\tilde \partial $}}}
\def\Lslash{\not{\hbox{\kern-2pt L}}}
\def\Qslash{\vert{\hbox{\kern-5pt Q}}}
\def\Rslash{\vert{\hbox{\kern-5.5pt R}}}

\def\mycomm#1{\nextline\strut\kern-3em{\tt ====> #1}\nextline}

\def\gray{\special{ps: 0.4 setgray}}
\def\black{\special{ps: 0.0 setgray}}

\newcommand{\draft}{
\newcount\timecount
\newcount\hours \newcount\minutes  \newcount\temp \newcount\pmhours

\hours = \time
\divide\hours by 60
\temp = \hours
\multiply\temp by 60
\minutes = \time
\advance\minutes by -\temp
\def\hour{\the\hours}
\def\minute{\ifnum\minutes<10 0\the\minutes
            \else\the\minutes\fi}
\def\clock{
\ifnum\hours=0 12:\minute\ AM
\else\ifnum\hours<12 \hour:\minute\ AM
      \else\ifnum\hours=12 12:\minute\ PM
            \else\ifnum\hours>12
                 \pmhours=\hours
                 \advance\pmhours by -12
                 \the\pmhours:\minute\ PM
                 \fi
            \fi
      \fi
\fi
}
\def\fullclock{\hour:\minute}
\gray
\font\Hugett  =cmtt12 scaled\magstep2
{\Hugett \strut \kern-3em Draft: \today,\clock}
\black
} 
\catcode`\@=11 
\def\lsim{\mathrel{\mathpalette\@versim<}}
\def\gsim{\mathrel{\mathpalette\@versim>}}
\def\@versim#1#2{\vcenter{\offinterlineskip
        \ialign{$\m@th#1\hfil##\hfil$\crcr#2\crcr\sim\crcr } }}
\catcode`\@=12 

\def\nextline{\hfill\break}

\def\mycomm#1{\nextline\strut\kern-6em{\tt ====> \ #1}\nextline}
\def\mpcomm#1{\nextline\strut\kern-6em{\tt MP COMMENT => \ #1}\nextline}
\def\nextline{\hfill\break}
\newcommand{\beq}{\begin{equation}}
\newcommand{\eeq}{\end{equation}}
\newcommand{\bea}{\begin{eqnarray}}
\newcommand{\eea}{\end{eqnarray}}
\newcommand{\Seff}{S_{\kern-0.1em\hbox{\small \it eff}}}
\newcommand{\Leff}{{\cal L}_{\kern-0.1em\hbox{\small \it eff}}}
\newcommand{\SMB}{S_{\kern-0.1em\hbox{\small \it MB}}}
\newcommand{\Smb}{S_{\kern-0.1em\hbox{\small \it m-b}}}

\def\eff{\hbox{\small\it eff}\,}
\def\SeffU{S_{\kern -0.1em \eff}[u]}

\def\eqref#1{(\ref{#1})}

\begin{document}
\begin{flushright}
CERN-PH-TH/2005-017\\
WIS-22-FEB-DPP\\
\end{flushright}
\vskip1.0cm
\begin{center}
{\Large\bf Exotic Baryons in Two-Dimensional QCD}
\end{center}
\vskip1cm
\begin{center}
{\bf John Ellis}\footnote{\tt john.ellis@cern.ch}\\
{\em Theory Division, CERN, Geneva, Switzerland}\\
\vspace*{0.5cm}
{\bf Yitzhak Frishman}\footnote{\tt yitzhak.frishman@weizmann.ac.il}\\
{\em  Department of Particle Physics\\
Weizmann Institute of Science, Rehovot, Israel}\\
\end{center}
\vskip0.5cm

%
\renewcommand{\baselinestretch}{0.95}
\centerline{\bf Abstract}
\vspace*{5mm}
\noindent

Two-dimensional QCD has often been used as a laboratory for studying the
full four-dimensional theory, providing, for example, an explicit
realization of baryons as solitons. We review aspects of
conventional baryons in two-dimensional QCD, including the classical and
quantum contributions to their masses. We then discuss the spectrum of
exotic baryons in two-dimensional QCD, commenting on the solitonic radius
inferred from the excitation spectrum as well as the two-dimensional
version of the Goldberger-Treiman relation relating meson couplings to
current matrix elements. Two-dimensional QCD provides strong overall
support to the chiral-soliton picture for the structure of normal and
exotic baryons in four dimensions.

\vspace*{0.5cm}
\begin{flushleft} CERN-PH-TH/2005-017\\
February 2005
\end{flushleft}


\vfill\eject

\section{Introduction}

Interest in baryons with `exotic' quantum numbers - that cannot be
composed of just three quarks - has been greatly stimulated recently by
several reports of baryons whose composition must include at least four
quarks and an antiquark - the so-called `pentaquark' states. The first of
these was the $\Theta^+ (1530)$, with ${\bar s} u d u d$ quantum
numbers~\cite{Theta}, and others include the $\Xi^{--} (1860)$~\cite{Xi}
and the $\Theta_c (3100)$~\cite{Thetac}. The existence of even the
$\Theta^+$ cannot yet be regarded as confirmed, and the other states have
each been seen in just one experiment. Nevertheless, reports of their
existence have stimulated considerable new debate about baryon structure.

The existence of such `exotic' baryons has long been predicted~\cite{G} by
chiral-soliton models~\cite{Skyrme}, and a very specific prediction of the
mass of the $\Theta^+$ was made~\cite{DPP} years before its apparent
confirmation.  However, the coincidence between the predicted and observed
masses may be just that, since there is still considerable uncertainty in
the chiral-soliton estimates~\cite{EKP}, and competing quark models have
been proposed~\cite{KL,JW} subsequent to the initial `pentaquark' reports.

Two-dimensional QCD has long been considered a useful theoretical
laboratory for studying non-perturbative strong-interaction issues such as
confinement and the large-$N_c$ expansion~\cite{'tHooft},
deep-inelastic~\cite{E} and high-energy scattering~\cite{BESW}, as well as
baryon structure~\cite{FS}. Clearly there are important differences
between QCD in four and two dimensions, however. One example is provided
by chiral symmetry and its breaking, on which we comment later in this
paper. Nevertheless, two-dimensional QCD may provide interesting insights
into the four-dimensional world.

The most convenient formulation for discussing baryons in two-dimension-
-al QCD is in terms of bosonic meson fields. This bosonization is exact,
and can be used to discuss baryons quantitatively as well as
qualitatively, in the large-$N_c$ limit. Bosonization methods and the
non-exotic baryon spectrum and properties are reviewed in~\cite{FS}.
As we recall in the next Section, baryons appear explicitly as solitons
made out of the meson fields, with masses that are a factor ${\cal
O}(N_c)$ larger. However, there are some important differences from QCD in
four dimensions. Although the mass of the lowest meson vanishes as the
quark mass $m_q \to 0$, it cannot be a Goldstone boson in two dimensions:
instead, it decouples as $m_q \to 0$. Likewise, the baryon mass also
vanishes in this limit. Its radius tends to a finite value in the
large-$N_c$ limit, which is of order $(m_q e_c)^{-1/2}$. As we discuss
below, the baryon-meson coupling also vanishes in this limit, but in such
a manner that the two-dimensional analogue of the Goldberger-Treiman
relation is valid up to corrections that are ${\cal O} (m_q/e_c)$.

Since there are no spin degrees of freedom in two dimensions, the
lowest-lying `three-quark' baryons correspond to the purely symmetric
Young tableau, and form a $\mathbf {10}$ representation of flavour SU(3),
analogous to the $\Delta$ multiplet of four-dimensional QCD. The ${\cal
O}(N_c)$ classical soliton mass can be calculated exactly, as can the
first quantum correction, which is explicitly ${\cal O}(N_c^0)$. This
provides an explicit formula for the two-dimensional analogue of the
baryon moment of inertia, which is model-dependent in four-dimensional
QCD.

As we discuss in Section 3, the next excited state is a `pentaquark' state
with the quantum numbers of four quarks and an antiquark which, because of
the quark wave-function symmetrization needed in two dimensions, forms a
$\mathbf {35}$ representation of flavour SU(3). In four-dimensional
chiral-soliton models, this multiplet is related to the ${\mathbf
{\overline {10}}}$ that is thought to contain the $\Theta^+$ and
$\Xi^{--}$ states, as well as a $\mathbf {27}$ representation, which have
no analogues in two dimensions.  The explicit mass formula for the
$\mathbf {35}$ state in two dimensions is very similar to the general
chiral-soliton formula in four dimensions~\cite{D}, with the differences
that there is no spin-dependent term $\propto J (J + 1)$ and the single
known soliton `moment of inertia' appears.

One may continue to discuss higher-lying baryons which would require
additional quark-antiquark pairs. The only allowed `heptaquark' forms a
totally symmetric $\mathbf {81}$ representation of SU(3), which is one of
those appearing in four dimensions at this level of exoticity. In the
large-$N_c$ limit, the extra energy required to progress to the next level
of exoticity: $qqq \to {\bar q} qqqq \to {\bar q} {\bar q} qqqqq \to
\cdots$ is explicitly ${\cal O}(N_c^0)$, as previously predicted on the
basis of four-dimensional chiral-soliton studies~\cite{D}.

We find that two-dimensional QCD provides strong overall support for the
chiral-soliton picture for baryon structure~\cite{Skyrme}.

\section{Review of Conventional Baryons in Two-Dimensional QCD}

\subsection{The Effective Action}

The natural form of the QCD$_2$ action is written in terms of gauge fields 
$A_\mu$ and fundamental quark fields $\Psi$:
\begin{equation}
S_F[\Psi,A_\mu]=\int d^2x\{-{1\over2e_c^2}Tr(F_{\mu\nu}
F^{\mu\nu})-\bar \Psi^{ai} {[(i\dslash +\Aslash) \Psi_i]}_a
+ m_q\bar \Psi^{ai} \Psi_{ai} \} ,
\label{action}
\end{equation}
where  $e_c$ is the quark coupling to the gauge fields, which has the 
dimension of a mass in $1+1$-dimensional space-time, $m_q$ is the common 
quark mass (we do not consider mass splittings in this paper), $a$ is the 
color index and $i$ the flavor index, and
\begin{equation}
F_{\mu\nu} \equiv \partial_\mu A_\nu -\partial_\nu A_\mu +i[A_\mu 
,A_\nu ]
\label{F}
\end{equation}
is the gauge field strength.

In two dimensions, one may go over to a completely bosonic description,
which is exact and particularly convenient for the discussion of the
baryon spectrum. As discussed in~\cite{FS}, various bosonization schemes 
are available. Here we use the $U(N_F\times N_c)$ scheme, in which the 
QCD$_2$ action is rewritten as
$$S[u,A_+,A_-] = S[u] - {1\over2e_c^{'2}}\int d^2x Tr(F_{\mu \nu}F^{\mu 
\nu})
 + $$
$$ {i\over2\pi}
\int d^2xTr(A_+u\partial_-u^\dagger+A_-u^\dagger\partial_+u) - $$
\begin{equation}
{1\over2\pi}\int d^2xTr(A_+uA_-u^\dagger-A_-A_+) + m^{'2} N_{\tilde m}
\int d^2x Tr(u+u^\dagger),
\label{uaction}
\end{equation}
where $u$ is a bosonic $U(N_F\times N_c)$ matrix, $S[u]$ is the 
Wess-Zumino-Witten action:
\begin{equation}
S[u]={1\over 8\pi}\int d^2xTr(\partial_\mu u\partial^\mu u^{-1})
+ {1\over12\pi}\int_B d^3y\varepsilon^{ijk}Tr(u^{-1}\partial_iu)
(u^{-1}\partial_ju)(u^{-1}\partial_ku),
\label{WZW}
\end{equation}
$e_c^{'} \equiv {\sqrt N_F}e_c$ and $m^{'2} \equiv m_q c \tilde 
m $, where $\tilde m$ is the normal-ordering scale, which can be determined by 
convenience. The constant $c = {1\over 2} e^\gamma$, where $\gamma$ is 
the Euler constant, so that $ c \approx 0.891$.

Gauging the $SU(N_c)$ subgroup and choosing the gauge $A_-=0$, the QCD$_2$ 
action becomes
$$ S[u,A_+] = S[u] +{1\over e_c^{'2}}\int d^2x Tr(\partial_-A_+)^2 + $$
\begin{equation}
{i\over 2\pi } \int d^2x Tr (A_+u\partial_-u^\dagger ) + 
m^{'2} N_{\tilde m} \int d^2x Tr(u+u^\dagger),
\label{newaction}
\end{equation}
where $N_{\tilde m}$ denotes normal ordering with respect to the scale 
$\tilde m$.

In this gauge, the action is quadratic in the gauge potentials, which may 
therefore be integrated out. Taking also the strong-coupling limit, 
in which the gauge coupling $e_c$ is much larger than the quark mass 
$m_q$, we can 
eliminate the color degrees of freedom entirely. Thus, we obtain an 
effective 
action expressed in terms of flavor degrees of freedom only:
\begin{equation}
S_{eff}[g]
=N_cS[g]+m^2N_m\int d^2x Tr_F(g+g^{\dagger}),
\label{Baction}
\end{equation}
where $g$ is a matrix representing $U(N_F)$, and the effective mass scale 
$m$ is 
\begin{equation}
m =[N_c c m_q({e_c\sqrt{N_F}\over \sqrt{2\pi}})^{\Delta_{c}}
]^{1\over 1+\Delta_{c}},
\label{mass}
\end{equation}
where the exponent
\begin{equation}
\Delta_c = {N_c^2-1\over N_c(N_c+N_F)}.
\label{Deltac}
\end{equation}
We recognize (\ref{Baction}) as an action of the type considered by Skyrme 
in QCD$_4$ to discuss baryonic solitons~\cite{Skyrme}. However, in QCD$_2$ 
it is the quark-mass term that plays the 
role of the stabilizing term, rather than the model-dependent 
higher-order terms used in QCD$_4$. Moreover, we emphasize that the above 
action is exact in QCD$_2$ in the strong-coupling (or small quark 
mass) limit.

In the large $N_c$ limit, which we use below to justify the semi-classical
approximation, the scale $m$ tends to a constant times $\sqrt {N_c e_c
m_q}$, where the constant is $0.56 N_F ^{1\over 4}$, which takes the value
0.74 for three flavors.

Note that we first take the strong-coupling limit $e_c \gg m_q$, and then
take $N_c$ to be large. This is different from the 't Hooft limit
~\cite{'tHooft}, where $e_c^2 N_c$ is held fixed.

\subsection{The Classical Soliton and its Mass}

In the spirit of the Skyrme model~\cite{Skyrme}, we first examine 
classical soliton
solutions of the bosonic action, which are heavy in the large-$N_c$ limit.  
We examine later the quantum corrections, using the semi-classical
approximation, and verify that they are small in the large-$N_c$ limit,
justifying {\it a posteriori} the assumption of a static, time-independent
first step.

Without loss of generality, we may assume for the lowest-energy state a 
diagonal form of the matrix $g(x)$:
\begin{equation}
g(x) = \left( e^{-i\sqrt{{4\pi\over N_c}}
\varphi_1},...,
e^{-i\sqrt{{4\pi\over N_c}}\varphi_{N_F}}\right).
\label{g}
\end{equation}
Using this Ansatz and redefining the constant term, the action density 
reduces to
\begin{equation}
\tilde S_d[g] = -\int dx\sum^{N_F}_{i=1} \left[
{1\over 2}
({d\varphi_i \over dx})^2
-2m^2 \left( cos \sqrt{{4\pi\over N_c}} \varphi_i -1
\right) \right],
\label{SG}
\end{equation}
which is a sum of standard sine-Gordon actions.
For each $\varphi_i$, the well-known solutions of the associated equations 
of motion are
\begin{equation}
\varphi_i (x)  = \sqrt{{4N_c \over \pi}}arctg
[ e^{(\sqrt{{8\pi\over N_c}}mx)} ],
\label{varphi}
\end{equation}
with the corresponding classical energy:
\begin{equation}
E_i = 4m\sqrt{{2N_c\over \pi}} \qquad i=1,....N_F .
\label{classen}
\end{equation}
Clearly the minimum energy configuration for this class of Ansatz
is when only one $\varphi_i$ is nonzero, for example
\begin{equation}
g_\circ(x) = Diag(1,1,....,e^{-i\sqrt{4\pi\over N_c} \varphi(x)}) .
\label{gcirc}
\end{equation}
We interpret this state as the lowest-lying baryon~\cite{FS}.

In the large-$N_c$ limit, the mass of this baryonic soliton is
\begin{equation}
M(classical) \approx 1.90 N_F ^{1\over 4} {\sqrt {e_c m_q}} N_c ,
\label{Mclass}
\end{equation}
In the particularly interesting case of three flavors, the coefficient 
becomes 2.50. We note that the mass is proportional to $N_c$, as we would 
expect. The proportionality to $\sqrt {e_c m_q}$ results from the form 
of the stabilizing term in (\ref{Baction}), and is specific to two 
dimensions.

As in the four-dimensional case, this classical soliton solution has 
non-zero baryon number as well as $Y$ charge, which is hypercharge 
normalized as a generator, so that the trace of $Y^2$ is 1/2:
\begin{equation}
Q^\circ _B = N_c \qquad Q^\circ _Y = -\sqrt{{(N_F-1)\over {2 N_F }}}N_c.
\label{Q}
\end{equation}
This observation follows from an explicit calculation of the charge 
densities of the 
soliton solutions, followed by their integration over space. We 
choose a normalization in which the quarks have baryon number $Q_B^\circ = 
1$, so the soliton has baryon number $N_c$. This soliton 
thus represents a physical baryon~\cite{FS}.

\subsection{Quantum Corrections and Allowed Representations}

We now calculate the quantum corrections. To this end, we allow the 
soliton to
rotate in $SU(N_F)$ space by a time-dependent amount $A(t)$:
\begin{equation}
g(x,t)=A(t) g_\circ(x) A^{-1}(t) \qquad A(t) \in U(N_F) ,
\label{rotate}
\end{equation}
We note that rotations by constant amounts are zero modes, and that 
charges of $SU(N_F)$, other than the $Y$ charge, now appear as results of 
such
rotations. It is straightforward to derive the effective action for $A(t)$.
The result is an extra contribution to the Hamiltonian from these quantum 
oscillations, as well as a constraint on the allowed physical states, the
latter coming from the Wess-Zumino term.
After a lengthy calculation~\cite{FS}, we obtain for the mass
\begin{equation}
E=M(classical)\{1 + ({\pi \over 2 N_c})^2 \left[C_2 (R) - N_c^2 
{(N_F-1)\over 2N_F} \right] \} ,
\label{qmass}
\end{equation}
where $M(classical)$ was given in (\ref{Mclass}) and $C_2 (R)$ is the 
value of the quadratic Casimir for the flavor
representation $R$ of the baryon.

In order to determine the allowed quantum states, the first constraint is 
that of baryon number, which 
should be $N_c$. The second is that the representation $R$ contains a 
state with the following value of the $Y$ charge:
\begin{equation}
\bar Q_Y= \sqrt{1\over 2(N_F-1)N_F}N_c .
\label{QY}
\end{equation}
All other states will be generated by applying the appropriate $SU(N_F)$ 
transformations to this one. Considering first states with only quarks and 
no antiquarks, the requirement that $Q_B=N_c$ implies that only
representations described by Young
tableaux with $N_c$ boxes appear. The extra constraint that
$Q_Y=\bar Q_Y$ implies that all $N_c$ quarks
are from $SU(N_F-1)$, not involving the $N_F$'th quark flavor.
These constraints are automatically obeyed in the totally symmetric 
representation of $N_c$ boxes, which is, in fact, the only 
representation possible in two dimensions. This is because
the states have to be constructed out of the
components of one complex vector $z$ as $\prod_{i=1}^{N_F}z_i^{n_i}$ with
$\sum_i n_i = N_c$. In four dimensions there is also the possibility of 
some mixed representations that are not totally symmetric, such as the 
octet in the case of three colors and three flavors, as one can 
obtain the necessary total symmetry of the non-color part of the wave 
function by combining mixed-symmetry representations of flavor and spin.
However, in QCD$_2$ for $N_c=3, N_F=3$ we get only the $\mathbf{10}$ of 
$SU(3)$.

Since the form of the stabilizing term is fixed, as well as the leading 
quadratic term in the action and the Wess-Zumino term, the leading quantum 
correction to the mass is also known exactly. Combining this with the 
classical term, the mass of the $\mathbf{10}$ baryon becomes
\begin{equation}
M(baryon)=M(classical)\left[1 + {\pi^2 \over 8}{(N_F-1)\over N_c} 
\right] ,
\label{tenmass}
\end{equation}
where $M(classical)$ was given in (\ref{Mclass}). We note that the quantum 
correction is indeed suppressed by a factor of $N_c$ as compared to
the classical term, as expected in any number of dimensions. On the other, 
hand, numerically the quantum correction $\sim 0.82$ for $N_c = 3, N_F = 
3$, which is not very small.

\subsection{Vibrational Modes}

The only static solutions of $QCD_2$ in the strong-coupling limit, are the
solitons we discussed above. As just discussed, their quantum corrections
are obtained by time-dependent rotations in flavor space, which are
suppressed by a factor of $N_c$ compared to the classical contribution to
the baryon mass.

To look for higher excitations that appear as vibrational modes, one
has first to look for time-dependent classical solutions.
Taking again the strong coupling limit, and assuming a diagonal 
form for $g$, we find the following equation:
\begin{equation} 
\partial_{tt}{\varphi}-\partial_{xx}{\varphi}+2 m^2 \sqrt{4\pi\over 
N_c}
sin \left(\sqrt{4\pi\over N_c}\varphi\right)=0,
\end{equation}
where $\varphi$ is one of the $\varphi_i$, in the notation of section (2.2).
We see that, on dimensional grounds, $\varphi$ is proportional to 
$\sqrt N_c$ times a function
of $m.x\over {\sqrt N_c}$ and  $m.t\over {\sqrt N_c}$.

It turns out that the only classical solution to this equation in the
sector with baryon number $B = N_c$ is the lightest baryon we found above
in the static case ~\cite{zz}. Thus, looking at the more general
time-dependent case does not give any new single-baryon states. However,
the time-dependent equation does give new meson states, as is well known
in Sine-Gordon theory. The low-lying mesons have masses of order
$\sqrt{m_q e_c}$, while the higher ones have masses of order $N_c$ times
that. We recall that the baryon masses are of the latter scale too, but
even the highest meson is lighter than twice the lightest baryon mass.

In a confining theory like QCD$_2$, we would expect an infinite tower of 
mesons, rather than the finite set mentioned above. This restriction is 
due to our strong-coupling limit, where we keep only the lowest mesons 
whose masses $\sim \sqrt{m_q e_c}$, for fixed $N_c$. All the other mesons
have masses $\sim e_c$, and so are infinitely heavier when $e_c / m_q$
goes to infinity. In the limit studied by 't Hooft~\cite{'tHooft}, 
in which $N_c$ is large with 
$e_c^2 N_c$ fixed, such an infinite tower does indeed appear, and
the mesons have squared masses
\begin{equation}
M^2_n \; \sim \; \left( {{e_c^2 N_c} \over\pi} \right) \pi^2 n
\label{tHooft}
\end{equation}
for large $n$.

\section{Generalization to Exotic Baryons}

\subsection{The First Exotic Baryon}

We now consider the case of the first exotic baryon ${\cal{E}}_1$, namely 
a state containing just one antiquark, which must also contain $N_c+1$ 
quarks. In two dimensions, for the reasons discussed earlier, the allowed 
state must be totally symmetric in the quarks.
For $N_c=3, N_F=3$, this state will be a $\mathbf{35}$ of flavor, to be 
compared with the $\mathbf{\overline{10}}$, $\mathbf{27}$ and 
$\mathbf{35}$ expected in four dimensions. The mass of this first 
exotic is easily found to be
\begin{equation}
M({\cal{E}}_1)=M(classical)\left[1 + {\pi^2\over 8}{1\over N_c}(3+N_F-
{6\over N_F})
+{3 \pi^2 \over 8}{1\over N_c^2} (N_F - {3\over N_F})\right] ,
\label{ME1}
\end{equation}
which is similar to the corresponding formula in~\cite{D}.
In the interesting case $N_c=3, N_F=3$, the ratio of the mass of the 
first exotic to that of the lightest baryon is
\begin{equation}
\frac{M_{\mathbf{35}}}{M_{\mathbf{10}}} \, = \, 
\frac{1+{\pi^2 \over 4}}{1 + {\pi^2\over 12}} .
\label{massratio}
\end{equation}
Numerically, this ratio is about 1.90. However, in this case the 
semi-classical approximation may not be a good 
approximation, as the quantum correction to the mass of 
the ${\cal{E}}_1$ state is larger than the classical term for $N_c = 3, 
N_F=3$.

Here we are more concerned with the concepts and principles of the soliton
model for baryons. On the other hand, we also note that the ratio of the
experimental masses of the $\Theta^+ (1530)$ and the nucleon is 1.63, so
the ratio in the QCD$_2$ model is only $\sim 10$~\% larger. However, this
could be an accident, since the QCD$_2$ calculation does not take into
account the fact that $m_s \ne m_{u,d}$ and the spin degree of freedom
that is important in QCD$_4$.

\subsection{Higher-Lying Exotic Baryons}

We now consider higher-lying exotics containing $p$ antiquarks and $N_c
+p$ quarks, which we call ${\cal{E}}_p$ baryons.
In two dimensions, according to the symmetry arguments 
already given above, the allowed states are
totally symmetric in the quarks, and also totally symmetric in the 
antiquarks. In the standard case $N_c=3, N_F=3$, the only allowed 
${\cal{E}}_2$ state is a $\mathbf{81}$ representation of flavor.
For general $N_c$ and $p$, considering the case $N_F=3$ as an example,
the mass of the ${\cal{E}}_p$ state is
\begin{equation}
M({\cal{E}}_p)=M(classical)\left[1 + {\pi^2\over {4 N_c^2}}[N_c(p+1) + 
p(p+2)] \right] ,
\label{Ep}
\end{equation}
corresponding to a correction that is considerably larger than unity. In 
this case, we would hesitate 
to advocate the semi-classical approximation for $N_c = 3$.

Nevertheless, we note that the spacing $\Delta$ between the 
${\cal{E}}_{(p+1)}$ and ${\cal{E}}_p$ exotic states, for large $N_c$, 
behaves like 
\begin{equation}
\Delta ={\pi^2 \over 4} {M(classical)\over N_c} .
\label{DeltaM}
\end{equation}
This amount becomes independent of $N_c$, and hence a constant 
additional mass, as the 
exoticity $p$ is increased. We recall that meson masses are also 
${\cal{O}}(N_c^0)$ for large $N_C$. This similarity supports the 
interpretation of this spacing as being analogous to the addition of a 
meson, which is indeed a quark-antiquark pair~\cite{D}.

Finally, we note that all the exotic baryons ${\cal E}_p$ with fixed
exoticity $p \ll N_c$ have masses much smaller than those of the
vibrational modes. Moreover, members of the ${\cal E}_p$ could mix with
non-exotic $\mathbf{10}$ baryons only via $SU(N_F)$-breaking mass effects,
and even this would be impossible for baryons with explicitly exotic
combinations of the charge and hypercharge quantum numbers.

\subsection{The Size of the QCD$_2$ Baryons}

In four dimensions, the QCD soliton has two moments of inertia, which
depend on the types of higher-order stabilizing terms used, and hence are
model-dependent. As already noted, in QCD$_2$, the quantum correction to
the mass depends on just one analogue of the moment of inertia. Its
classical expression is $\int dr r^2 \rho(r)$, where $\rho (r) $ is the
contribution to the classical soliton mass from a shell of radius $r$,
$M(classical) = \int dr \rho (r)$. Hence, we can define an effective 
soliton radius by
\begin{equation}
\ll r \gg \, \equiv \, \sqrt{<r^2>}, \; {\rm where} \; <r^2> \, = \, 
\frac{I}{M{(classical)}}.
\label{radius}
\end{equation}
Comparing the quantum mass formula (\ref{Ep}) with the corresponding 
formula in QCD$_4$ in the limit $N_c \gg p \gg 1$, we infer that
\begin{equation}
I \, = \, \frac{N^2_c}{\pi^2 M(classical)},
\label{I}
\end{equation}
and hence that
\begin{equation}
\ll r \gg \, = \, \sqrt{\frac{I}{M(classical)}} \, = \, \frac{N_c}{\pi 
M(classical)} \, = \, \frac{1}{1.90 \pi N_F^{1/4} \sqrt{e_cm_q}}.
\label{r}
\end{equation}
We see explicitly that $\ll r \gg = {\cal{O}}(N_c^0)$, as expected also in 
four dimensions. The fact that $\ll r \gg$ depends on $m_q$ reflects the 
special 
feature of the stabilizing term in QCD$_2$, and has no implications for 
four dimensions.

As a curiosity, note that, if we take $n_F = 3$ flavors, $e_c = 100$~MeV 
for the coupling and $m_q = 10$~MeV for the quark mass, we obtain an 
effective baryon radius $\approx 1 / (248~{\rm Mev})$.

\subsection{The Goldberger-Treiman Relation in QCD$_2$}

It is well known that a continuous symmetry cannot be broken spontaneously 
in two
dimensions, and hence there can be no Goldstone bosons. On the other hand,
the mass of the lowest-lying QCD$_2$ meson vanishes in the limit $m_q \to 
0$,
just like that of the four-dimensional pion. 
In two dimensions, however, the massless ``pion'' decouples, whilst in 
four dimensions
it has has non-vanishing
couplings in the limit $m_q \to 0$, which are proportional to
axial-current matrix elements according to generalized Goldberger-Treiman
relations:
\begin{equation}
g_{\pi N N} \, = \, \frac{2 M_N}{f_\pi} g_A ,
\label{GT}
\end{equation}
where (omitting irrelevant Lorentz factors) $f_\pi$ is the coupling of the 
four-dimensional $\pi$ meson to the 
axial current ${\cal A}$, and $g_A$ is the matrix element of the axial 
current ${\cal A}$ in the 
nucleon $N$. In QCD$_2$, the flavor axial current ${\cal A}$ is 
directly related to the corresponding vector current ${\cal V}$:
\begin{equation}
{\cal A}_\mu \, = \, \epsilon_{\mu \nu} {\cal V}^\nu .
\label{dual}
\end{equation}
Hence, $f_\pi$ and $g_A$ are directly related to the flavor quantum 
numbers of the lowest-lying meson and baryon, respectively, and thus
non-vanishing in the limit $m_q \to 0$.
In QCD$_2$, the nucleon mass $\to 0$ as $m_q \to 0$, as we 
have already seen in (\ref{Mclass}), and so also do $g_{\pi N N}$ and the 
other $\pi$-baryon couplings. However, 
the ratio $g_{\pi N N} \over {2 M_N}$ is non-zero in this limit, being 
equal to $ g_A \over {f_\pi}$, as in the Goldberger-Treiman relation (\ref{GT}).


We firmly expect the corresponding generalized Goldberger-Treiman 
relations to hold for the couplings of light pseudoscalar mesons 
to baryonic solitons in QCD$_4$.

Finally, we note a corollary of the above remarks for the ${\mathbf{35}} -
{\mathbf{10}}$ - meson coupling or, more generally, for meson couplings
between baryons with differing degrees of exoticity. Such couplings would
be related by generalized Goldberger-Treiman relations to matrix elements
of axial currents ${\cal A}$ between baryons in different representations
of $SU(N_F)$, which are in turn related by (\ref{dual}) to off-diagonal 
matrix
elements of vector currents ${\cal V}$. Since these must vanish, so also
do the meson couplings between baryons with differing degrees of
exoticity, at least in the large-$N_c$ limit in which our approximation to
QCD$_2$ is valid. This parallels the observation that the $K - N -
\Theta^+$ vanishes in the large-$N_c$ limit in QCD$_4$~\cite{P}.

How large may the ${\mathbf{35}} - {\mathbf{10}}$ - meson coupling be?
Since it vanishes to leading order in $N_c$, we expect it to go to a
constant in the large-$N_c$ limit. A generic matrix element of the axial
current between two baryon states has the dimensionality of a mass. In the
large-coupling limit, the only relevant mass parameter is $\sqrt {e_c
m_q}$, as we have seen above. Therefore, we expect the matrix element also
to be proportional to $\sqrt {e_c m_q}$. This is consistent with its
vanishing for zero quark mass, which we expect on the basis of the general
arguments given above. There is no possibility of compensating for the
large-$N_c$ suppression by a different dependence on $e_c$.

\section{Summary}

QCD$_2$ provides a laboratory where the approximations made in deriving
the chiral-soliton model~\cite{Skyrme} are explicit, and the corrections
to the lowest-order calculations can be calculated exactly. In this paper,
we have extended previous studies of baryons in QCD$_2$~\cite{FS} to
include the analogues of the exotic baryons that have been predicted in
four-dimensional QCD~\cite{DPP}, which inspired experimental searches that
have produced some positive reports~\cite{Theta,Xi,Thetac}. Apart from
differences specific to the structure of QCD$_2$, baryons in the
two-dimensional model have many similarities to the expectations
formulated on the basis of chiral-soliton models in QCD$_4$~\cite{D}. We
have exhibited explicitly the natures of the lowest-lying exotic baryons
and calculated both their classical masses and the leading quantum
corrections.

We consider that this analysis provides considerable support to the 
four-dimensional chiral-soliton model~\cite{Skyrme}.

\section*{Acknowledgements}

Y. F. would like to thank the Theory Division at CERN for their hospitality
when part of this work was done.

\end{document}